\documentclass[prl,aps,showpacs,twocolumn,floatfix]{revtex4}
\usepackage{epsfig} \usepackage{graphics} \usepackage{bm}
\usepackage{amssymb}
\usepackage{graphicx}
\begin{document}

\preprint{Lebed-Rapids-LN}

\title{Four-fold anisotropy of the parallel upper critical magnetic
field in a pure layered d-wave superconductor at $T = 0$}

\author{A.G. Lebed$^*$ and O. Sepper}

\affiliation{Department of Physics, University of Arizona, 1118 E.
4-th Street, Tucson, AZ 85721, USA}

\begin{abstract}
It is well known that a four-fold symmetry of the parallel upper
critical magnetic field disappears in the Ginzburg-Landau (GL)
region in quasi-two-dimensional (Q2D) $d$-wave superconductors.
Therefore, it has been accurately calculated so far as a
correction to the GL results, which is valid close to
superconducting transition temperature and is expected to be
stronger at low temperatures. As to the case $T=0$, some
approximated methods have been used, which are good only for
closed electron orbits and unappropriate for the open orbits which
exist in a parallel magnetic field in Q2D superconductors. For the
first time, we accurately calculate the four-fold anisotropy of
the parallel upper critical magnetic field in a pure Q2D $d$-wave
superconductor at $T=0$, where it has the highest possible value.
Our results are applicable to Q2D $d$-wave high-Tc and organic
superconductors.
\end{abstract}


\maketitle

Since the discovery of unconventional $d$-wave superconductivity
in high-temperature superconductors [1], physical consequences of
$d$-wave electron pairing have been intensively investigated. One
of such physical properties is a four-fold symmetry of the
parallel upper critical magnetic field in these
quasi-two-dimensional (Q2D) superconductors [2-5]. From the
beginning, it was recognized that the four-fold anisotropy of the
parallel upper critical magnetic field disappears in the
Ginzburg-Landau (GL) region [6] and has to be calculated as a
non-local correction to the GL results [3,4]. Another approach was
calculation of the parallel upper critical magnetic field at low
temperatures and even at $T=0$ [2,7-9] using approximate method
[10], which was elaborated for unconventional superconductors with
closed electron orbits in an external magnetic field. Note that
Q2D conductors in a parallel magnetic field are characterized by
open electron orbits, which makes the calculations [2,7-9] to be
unappropriate.

The goal of our article is to suggest an appropriate method to
calculate the parallel upper critical magnetic field in a Q2D
$d$-wave superconductor. For this purpose, we explicitly take into
account almost cylindrical shape of its Fermi surface (FS) and the
existence of open electron orbits in a parallel magnetic field. We
use the Green's functions formalism to obtain the Gor'kov's gap
equation in the field. As an important example, we numerically
solve this integral equation to obtain the four-fold anisotropy of
the parallel upper critical magnetic field in a $d_{x^2-y^2}$-wave
Q2D superconductor with isotropic in-plane FS. In particular, we
demonstrate that the so-called supercondcting nuclei at $T=0$
oscillate in space in contrast to the previous results [2,7-9]. We
also suggest the gap equation which take both the orbital and
paramagnetic spin-splitting mechanisms against superconductivity.

Below, we consider a layered superconductor with the following Q2D
electron spectrum, which is an isotropic within the conducting
plane:
\begin{equation}
\epsilon({\bf p})= \epsilon (p_x,p_y) - 2 t_{\perp} \cos(p_z c^*),
\ \ \  t_{\perp} \ll \epsilon_F,
\end{equation}
where
\begin{equation}
\epsilon (p_x,p_y) = \frac{ (p^2_x + p^2_y)}{2m} \ , \ \ \
\epsilon_F = \frac{p^2_F}{2m} \ .
\end{equation}
[Here, $m$ is the effective in-plane electron mass, $t_{\perp}$ is
the integral of overlapping of electron wave functions in a
perpendicular to the conducting planes direction; $\epsilon_F$ and
$p_F$ are the Fermi energy and Fermi momentum, respectively;
$\hbar \equiv 1$.] The parallel magnetic field is assumed to be
applied along ${\bf x}$ axis,
\begin{equation}
{\bf H} = (H,0,0) \ ,
\end{equation}
where vector potential of the field is convenient to choose in the
form:
\begin{equation}
{\bf A} = (0,0,Hy) \ .
\end{equation}

Electron motion within the conducting plane is supposed to be free
(2), therefore, we can make the following substitutions in the
electron energy (1) and (2):
\begin{equation}
p_x \rightarrow - i \biggl( \frac{\partial }{ \partial x} \biggl),
\ \ p_y \rightarrow - i \biggl( \frac{\partial }{
\partial y} \biggl),
\end{equation}
whereas, for the perpendicular electron motion, we can perform the
so-called Peierls substitution:
\begin{equation}
p_z c^* \rightarrow p_z c^*- \biggl( \frac{\omega_c}{v_F} \biggl)
y, \ \ \omega_c = \frac{e v_F c^* H}{c}.
\end{equation}

As a result, the electron Hamiltonian in the magnetic field can be
represented as:
\begin{equation}
\hat H = - \frac{1}{2m} \biggl( \frac{\partial^2 }{ \partial x^2}
+ \frac{\partial^2 }{ \partial y^2} \biggl) - 2 t_{\perp} \cos
\biggl(p_z c^* - \frac{\omega_c}{v_F}y \biggl),
\end{equation}
where electron wave functions, for $\omega_c \ll \epsilon_F$, can
be written in the mixed $(p_x, y, p_z)$ representation:
\begin{eqnarray}
&\Psi^{\pm}_{\epsilon}(x,y,z) =  \ \exp( i p_x x) \ \exp[\pm i
p^0_y(p_x) y] \ \exp( i p_z z)
\nonumber\\
&\times \psi_{\epsilon}^{\pm}(p_x,y,p_z), \ \  p^0_y(p_x) =
\sqrt{p_F^2-p^2_x}.
\end{eqnarray}
We stress that for the main part of the Q2D Fermi surface (1),(2)
the following condition of quasiclassical motion is valid:
\begin{equation}
p^0_y(p_x) \sim p_F.
\end{equation}
It is easy to prove that, in this case, we can represent the
electron Hamiltonian (7) for the wave functions
$\psi_{\epsilon}^{\pm}(p_x,y,p_z)$ in Eq.(8) as
\begin{eqnarray}
&\biggl\{ \frac{1}{2m} \biggl[p^2_F \pm 2 i p^0_y(p_x)
\frac{d}{dy}\biggl] - 2 t_{\perp} \cos \biggl(p_z c^* -
\frac{\omega_c}{v_F} y \biggl) \biggl\}
\nonumber\\
&\times \psi_{\epsilon}^{\pm}(p_x,y,p_z) = (\epsilon +
\epsilon_F)\ \psi_{\epsilon}^{\pm}(p_x,y,p_z).
\end{eqnarray}
We point out that energy $\epsilon$ in Eq.(10) is counted from the
Fermi level, $\epsilon_F =p^2_F/2m$. Then, it is straightforward
to rewrite Eq.(10) in a more convenient way:
\begin{eqnarray}
&\biggl[ \pm i v^0_y(p_x) \frac{d}{dy} - 2 t_{\perp} \cos
\biggl(p_z c^* - \frac{\omega_c}{v_F} y \biggl) \biggl]
\psi_{\epsilon}^{\pm}(p_x,y,p_z)
\nonumber\\
&= \epsilon \ \psi_{\epsilon}^{\pm}(p_x,y,p_z), \ \ \ v^0_y(p_x) =
p^0_y(p_x)/m .
\end{eqnarray}

Note that Eq.(11) is very general. For instance, for a pure case,
$\pi T_c \gg 1/\tau_0$, it contains quantum effects of an electron
motion in a magnetic field in the Brillouin zone, where $T_c$ is a
superconducting temperature at $H=0$ and $\tau_0$ is scattering
time of electrons with impurities. Quantum nature of Eq.(11)
follows from periodicity of its solutions in the variable $y$,
which takes into account multiple electron reflections from
boundaries of the zone. As was shown before, such effects are
important at very high magnetic fields, $\omega_c(H_{c2}) \geq 4
t_{\perp}$, or at very low temperatures, $2 \pi T \leq
\omega_c(H_{c2})$, where stabilization and the reentrance of
superconducting phase is expected [11-13]. Below, we study the
opposite case - a pure superconductor, where electron motion
between the conducting planes is quasiclassical [14], and,
therefore, we can take into account only the first order terms
with respect to the magnetic field in Eq.(11) [14]. As a result,
we obtain, instead of Eq.(11):
\begin{eqnarray}
&\biggl[ \pm i v^0_y(p_x) \frac{d}{dy}  -2t_{\perp} \cos (p_zc^*)
-\biggl( \frac{2t_{\perp} \omega_c y}{v_F} \biggl) \sin (p_zc^*)
\nonumber\\
& -\mu_B \sigma H \biggl] \ \psi_{\epsilon}^{\pm}(p_x,y,p_z) =
\epsilon \ \psi_{\epsilon}^{\pm}(p_x,y,p_z).
\end{eqnarray}
[We point out that, in Eq.(12), we also take into account the
Pauli paramagnetic spin-splitting effects in the magnetic field,
where $\sigma = +(-)$ corresponds to electron spin up(down) in the
field, with $\mu_B$ being the Bohr magneton.] It is important that
Eq.(12) can be exactly solved:
\begin{eqnarray}
&\psi_{\epsilon}^{\pm}(p_x,y,p_z) = \exp \biggl[\mp i
\frac{\epsilon y}{v^0_y(p_x)} \bigg] \exp \biggl[\mp i \frac{2
t_{\perp}\cos (p_zc^*) y}{v^0_y(p_x)} \biggl]
\nonumber\\
&\times \exp \biggl[ \mp i \frac{t_{\perp} \omega_c
y^2}{v^0_y(p_x) v_F} \sin(p_z c^*) \biggl] \ \exp \biggl[\mp i
\frac{\mu_B \sigma H y}{v^0_y(p_x)} \bigg].
\end{eqnarray}

To find the Green's functions of non-interacting electrons in a
magnetic field in the mixed $(p_x,y,p_z)$ representation, we use
the following equation [15,16]:
\begin{eqnarray}
&\biggl[i \omega_n  \mp i v^0_y(p_x) \frac{d}{dy}  +2t_{\perp}
\cos (p_zc^*) +\biggl( \frac{2t_{\perp} \omega_c y}{v_F} \biggl)
\sin (p_zc^*)
\nonumber\\
& +\mu_B \sigma H \biggl] \ g_{i \omega_n}^{\pm}(p_x;y,y_1;p_z) =
\delta(y-y_1) \ ,
\end{eqnarray}
where $\omega_n$ is the so-called Matsubara frequency [15]. It is
possible to solve Eq.(14) analytically and find the following
expressions for the electron Green's functions:
\begin{eqnarray}
&g^{\pm}_{i \omega_n} (p_x; y, y_1; p_z) = - i \frac{ sgn(
\omega_n)}{v^0_y(p_x)}
 \exp \biggl[ \pm \frac{\omega_n (y-y_1)}{v^0_y(p_x)} \biggl]
\nonumber\\
&\times \exp \biggl[\mp i \frac{\mu_B \sigma H
(y-y_1)}{v^0_y(p_x)} \bigg] \exp \biggl[\mp i \frac{2
t_{\perp}\cos (p_zc^*) (y-y_1)}{v^0_y(p_x)} \biggl]
\nonumber\\
&\times \exp \biggl[ \mp i \frac{t_{\perp} \omega_c
(y^2-y^2_1)}{v^0_y(p_x) v_F} \sin(p_z c^*) \biggl] .
\end{eqnarray}

Using the known Green's functions (15), it is possible to derive
the so-called gap equation, determining the upper critical
magnetic field for unconventional superconductivity, by means of
the linearized Gor'kov's equation for the nonuniform
superconductivity (see equation (17.9) of Ref.[17]). Note that the
gap Eq.(16) can also be obtained as a quasi-classical limit in a
magnetic field of the master equation of our Ref.[13]:
\begin{eqnarray}
&\Delta(\phi, y) = \int_0^{2 \pi} \frac{d \phi_1}{2 \pi}
U(\phi,\phi_1) \int^{\infty}_{ |y-y_1| > d |\sin \phi_1|}
\nonumber\\
&\times \frac{2 \pi T dy_1}{v_F \sin \phi_1 \sinh \biggl( \frac{ 2
\pi T |y-y_1|}{ v_F \sin \phi_1} \biggl) } \cos \biggl[ \frac{2 k
\mu_B \sigma H (y-y_1)}{v^0_y(p_x)} \bigg]
\nonumber\\
&\times J_0 \biggl[ \frac{2 t_{\perp} \omega_c}{v^2_F \sin \phi_1}
(y^2-y^2_1)\biggl]   \ \Delta(\phi_1, y_1)\ ,
\end{eqnarray}
where  $d$ is the cut-off distance, $J_0(...)$ is the zero order
Bessel function. In Eq.(16) the superconducting gap
$\Delta(\phi,y)$ depends on a center of mass of the BCS pair, $y$,
and on the position on the cylindrical FS, where $\phi_1$ is the
polar angle, corresponding to two-component vector ${\bf p} =
[p_x, p^0_y(p_x)]$. Electron-electron interactions, $U(\phi
,\phi_1)$, depend only on in-plain momenta [e.g., $U(\phi,\phi_1)
= g$ for $s$-pairing, $U(\phi,\phi_1) = g \cos (\phi)
\cos(\phi_1)$ for $p$-pairing, and $U(\phi,\phi_1) = g \cos
(2\phi) \cos(2\phi_1)$ for $d_{x^2-y^2}$-pairing].

As follows from our derivation, Eq.(16) defines the upper critical
field in conventional and unconventional pure type II Q2D
superconductors. It is important that $k=1$ in Eq.(16) corresponds
to singlet $s$-wave and $d$-wave electron pairings, which takes
into account the Pauli paramagnetic spin-splitting effects against
superconductivity. Nevertheless, in some paramagnetically
insensitive triplet phases (for example, when ${\vec d}$-vector
[17] is perpendicular to the conducting planes) the parameter $k$
is equal to $0$ in Eq.(16). Note that, in this paper, we calculate
the maximal four-fold anisotropy of the orbital upper critical
magnetic field for a singlet $d$-wave superconductor, where the
paramagnetic term is small in Eq.(16). Therefore, we can rewrite
Eq.(16) in the following way:
\begin{eqnarray}
&\Delta(\phi, y) = \int_0^{2 \pi} \frac{d \phi_1}{2 \pi}
U(\phi,\phi_1) \int^{\infty}_{ |y-y_1| > d |\sin \phi_1|}
\nonumber\\
&\times \frac{2 \pi T dy_1}{v_F \sin \phi_1 \sinh \biggl( \frac{ 2
\pi T |y-y_1|}{ v_F \sin \phi_1} \biggl) }
\nonumber\\
&\times J_0 \biggl[ \frac{2 t_{\perp} \omega_c}{v^2_F \sin \phi_1}
(y^2-y^2_1)\biggl]   \ \Delta(\phi_1, y_1)\ .
\end{eqnarray}
To show that Eq.(17) does not have a singularity at $\phi_1=0$, we
introduce new variable of integration, $y_1 = z \sin \phi_1 +y$,
and rewrite Eq.(17) in the more convenient way:
\begin{eqnarray}
&\Delta(\phi, y) = \int_0^{2 \pi} \frac{d \phi_1}{2 \pi}
U(\phi,\phi_1) \int^{\infty}_{|z|>d} \frac{2 \pi T dz}{v_F \sinh
\biggl( \frac{ 2 \pi T z}{ v_F } \biggl)}
\nonumber\\
&\times J_0 \biggl\{ \frac{2 t_{\perp} \omega_c}{v^2_F} [z(2y+z
\sin \phi_1)] \biggl\}   \ \Delta(\phi_1, y + z \sin \phi_1)\ ,
\end{eqnarray}

Let us rotate in-plane magnetic field by polar angle $\alpha$,
where
\begin{equation}
{\bf H}= (H \cos \alpha, H \sin \alpha, 0) \ .
\end{equation}
In this case, it is possible to rewrite the gap equation Eq.(18)
in the following way:
\begin{eqnarray}
&\Delta_{\alpha}(\phi, y) = \int_0^{2 \pi} \frac{d \phi_1}{2 \pi}
U(\phi,\phi_1) \int^{\infty}_{|z|>d} \frac{2 \pi T dz}{v_F \sinh
\biggl( \frac{ 2 \pi T z}{ v_F } \biggl)}
\nonumber\\
&\times J_0 \biggl\{ \frac{2 t_{\perp} \omega_c}{v^2_F} [z(2y+z
\sin (\phi_1-\alpha))] \biggl\}
\nonumber\\
&\times\Delta_{\alpha}[\phi_1, y + z \sin (\phi_1-\alpha)]\ ,
\end{eqnarray}
which can be transformed into
\begin{eqnarray}
&\Delta_{\alpha}(\phi, y) = \int_0^{2 \pi} \frac{d \phi_1}{2 \pi}
U(\phi,\phi_1+\alpha) \int^{\infty}_{|z|>d} \frac{2 \pi T dz}{v_F
\sinh \biggl( \frac{ 2 \pi T z}{ v_F } \biggl)}
\nonumber\\
&\times J_0 \biggl\{ \frac{2 t_{\perp} \omega_c}{v^2_F} [z(2y+z
\sin \phi_1)] \biggl\}
\nonumber\\
&\times\Delta_{\alpha}(\phi_1+\alpha, y + z \sin \phi_1)\
\end{eqnarray}
by shifting the angle of integration in Eq.(20): $\phi_1
\rightarrow \phi_1 +\alpha$.

Let us consider a model $d_{x^2-y^2}$ electron superconducting
coupling in the traditional factorized form,
\begin{equation}
U(\phi,\phi_1)= g \cos(2 \phi) \cos(2 \phi_1),
\end{equation}
then, for the solution of the gap equation,
\begin{equation}
\Delta_{\alpha}(\phi,y)= \sqrt{2} \cos(2 \phi) \Delta_{\alpha}(y),
\end{equation}
it can be expressed as
\begin{eqnarray}
&\Delta_{\alpha}(y) = g \int^{\infty}_{d} \frac{2 \pi T dz}{v_F
\sinh \biggl( \frac{ 2 \pi T z}{ v_F } \biggl)}
\nonumber\\
&\biggl< J_0 \biggl\{ \frac{2 t_{\perp} \omega_c}{v^2_F} [z(2y+z
\sin \phi_1)] \biggl\} \Delta_{\alpha}(y + z \sin \phi_1)
\nonumber\\
&\times[1+\cos(4 \alpha) \cos(4 \phi_1)] \biggl>_{\phi_1}.
\end{eqnarray}

Here, we derive the GL equation for $d_{x^2-y^2}$-wave
superconductor in a parallel magnetic field to determine the GL
slope of the field. To this end, we expand the superconducting gap
and Bessel function in Eq.(24) with respect to small parameter, $z
\ll v_F/(\pi T_c)$:
\begin{eqnarray}
&&\Delta_{\alpha}(y+z \sin \phi_1) \approx \Delta_{\alpha}(y) +
\frac{1}{2} z^2 \sin^2 \phi_1 \biggl[ \frac{d^2
\Delta_{\alpha}(y)}{dy^2} \biggl] ,
\nonumber\\
 &&J_0 \biggl\{ \frac{2 t_{\perp}
\omega_c}{v^2_F} [z (z \sin \phi_1 +2y)] \biggl\} \approx
1-\frac{4 t^2_{\perp} \omega^2_c}{v^4_F} z^2 y^2.
\end{eqnarray}
Now we substitute the expansions (25) into the integral gap
Eq.(24) and average over angle $\phi_1$:
\begin{eqnarray}
&&-\frac{1}{4} \biggl[ \frac{d^2 \Delta_{\alpha}(y)}{dy^2} \biggl]
\int_0^{\infty} \frac{2 \pi T_c z^2 d z}{v_F \sinh \biggl( \frac{
2 \pi T_c z}{ v_F} \biggl)}
\nonumber\\
&&+ y^2 \Delta_{\alpha}(y) \frac{4t^2_{\perp}\omega^2_c}{v^4_F}
\int_0^{\infty} \frac{2 \pi T_c z^2 d z}{v_F \sinh \biggl( \frac{
2 \pi T_c z}{ v_F} \biggl)}
\nonumber\\
&&+\Delta_{\alpha}(y) \biggl[ \frac{1}{g} - \int_d^{\infty}
\frac{2 \pi T d z}{v_F \sinh \biggl( \frac{ 2 \pi T z}{ v_F}
\biggl)} \biggl]=0.
\end{eqnarray}
[Note that the average of the second contribution to Eq.(24),
which contains the angular dependence $\cos(4 \alpha)$, is zero.
Therefore, in the GL area the four-fold anisotropy of the parallel
upper critical field disappears.] At zero magnetic field, we have
the following equation, which determines the superconducting
transition temperature, $T_c$ :
\begin{equation}
 \frac{1}{g} = \int_d^{\infty}  \frac{2 \pi T_c d z}{v_F
\sinh \biggl( \frac{ 2 \pi T_c z}{ v_F} \biggl)} \ .
\end{equation}
As a result of transformations of Eqs.(26),(27), we obtain the
following GL differential equation,
\begin{equation}
- \xi^2_{\parallel} \biggl[ \frac{d^2 \Delta_{\alpha}(y)}{dy^2}
\biggl] + \biggl(\frac{2\pi H}{\phi_0}\biggl)^2 \xi^2_{\perp} y^2
\Delta_{\alpha}(y) -\tau \Delta_{\alpha}(y) = 0,
\end{equation}
where we introduce the parallel and perpendicular GL coherent
lengths
\begin{equation}
\xi_{\parallel} = \frac{\sqrt{7 \zeta(3)}v_F}{4 \sqrt{2} \pi T_c},
\ \ \ \xi_{\perp} = \frac{\sqrt{7 \zeta(3)} t_{\perp} c^*}{2
\sqrt{2} \pi T_c},
\end{equation}
and where we take into account that [18]:
\begin{equation}
\int^{\infty}_0 \frac{z^2 dz}{\sinh(z)}  = \frac{7}{3} \zeta(3) \
.
\end{equation}
[Note that, in Eqs.(28)-(30), $\zeta(x)$ is the Riemann
zeta-function, $\phi_0 = \frac{\pi c}{e}$ is the magnetic flux
quantum, $\tau=\frac{T_c-T}{T_c}$.]
 It is important that the GL
Eq.(28), which is similar to the Schr\"{o}dinger equation for a
harmonic oscillator, defines the parallel GL upper critical
magnetic field as the minimal field, where it has a solution. From
Eq.(28), it follows that the parallel upper critical magnetic
field is equal to:
\begin{equation}
H^{GL}_{c2}(T) = \tau \biggl( \frac{\phi_0}{2 \pi \xi_{\parallel}
\xi_{\perp}} \biggl) = \tau \biggl[ \frac{8 \pi^2 c T^2_c}{7
\zeta(3) e v_F t_{\perp} c^*} \biggl]
\end{equation}
and does not depend on the direction of magnetic field (19) (i.e,
the angle $\alpha$). [We note that the GL formula for the upper
critical magnetic field (31) is valid only if $\tau \ll 1$ and can
be also obtained from Ref.[19].]

The next our step is to define the four-fold anisotropy of the
parallel upper critical magnetic field in the $d_{x^2-y^2}$ Q2D
superconductor as a function of the angle $\alpha$ (19) at zero
temperature, where it takes a maximum value. To this end, we
rewrite the integral Eq.(24) for $T=0$ [20]:
\begin{eqnarray}
&\Delta_{\alpha}(y) = g \int^{\infty}_d \frac{dz}{z} \biggl< J_0
\biggl\{ \frac{2 t_{\perp} \omega_c}{v^2_F} [z(2y+z \sin \phi_1)]
\biggl\}
\nonumber\\
&\times[1+\cos(4 \alpha) \cos(4 \phi_1)] \Delta_{\alpha}(y + z
\sin \phi_1)\biggl>_{\phi_1} \ ,
\end{eqnarray}
and introduce new convenient for the further numerical solutions
of this equation variables:
\begin{equation}
\tilde z = \frac{\sqrt{2t_{\perp} \omega_c}}{v_F} z , \ \ \ \tilde
y = \frac{\sqrt{2t_{\perp} \omega_c}}{v_F} y .
\end{equation}
[Here, at $T=0$ we define the upper critical magnetic field in the
framework of the Landau theory of the second order phase
transitions, as it is done, for example, in isotropic 3D case in
Ref.[14]. Therefore, we disregard the possible existence of
quantum phase transitions.]$\\$
In new variables Eq.(32) can be
written as follows
\begin{eqnarray}
&\Delta_{\alpha}(\tilde y) = g \int^{\infty}_d \frac{d \tilde
z}{\tilde z} \biggl< J_0 [\tilde z(2 \tilde y + \tilde z \sin
\phi_1)]
\nonumber\\
&\times[1+\cos(4 \alpha) \cos(4 \phi_1)]\Delta_{\alpha}(\tilde y +
\tilde z \sin \phi_1) \biggl>_{\phi_1} \ .
\end{eqnarray}

Below, we solve integral Eq.(34) numerically. The typical example
of its solution for the superconducting nucleus, $\Delta(\tilde
y)$, is shown in Fig.1, where it oscillates and changes its sign
with the changing coordinate $\tilde y$. These oscillations are
consequences of the open nature of electron trajectories in a
parallel magnetic field for the Q2D electron spectrum (1),(2). As
we have shown, they result in the appearance of the oscillating
Besssel function in the gap Eq.(34). This typical behavior of the
superconducting nuclei is in a sharp contrast with the
calculations of the four-fold anisotropy at $T=0$, which were done
before [2,7-9]. The reason for that is the fact that the previous
calculations didn't take into account open nature of electron
orbits in Q2D conductors in a parallel magnetic field. Numerically
calculated from Eq.(34) angular dependence of the parallel upper
critical magnetic field of a Q2D $d_{x^2-y^2}$ superconductor is
shown in Fig.2, where there is a sharp peak at $\alpha=0$ and a
shallow minima at $\alpha=45^0$. The calculated magnitude of the
four-fold anisotropy is $[H_{c2}(0^0)-H_{c2}(45^0)]/H_{c2}(22.5^0)
= 0.13$, which is high than that reported before [2]. In addition,
from Fig.2, it is clear that the calculated by us anisotropic term
is not of a pure $\cos(4 \alpha)$ form as were stated in the all
previous calculations [2,7-9].

\begin{figure}[t]
\centering
\includegraphics[width=0.5\textwidth]{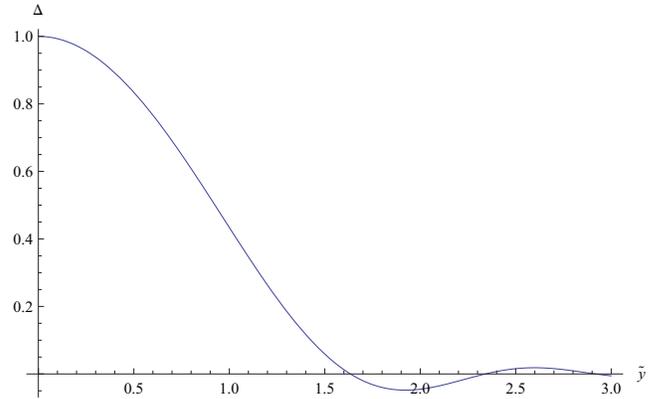}
\caption{Dependence of the superconducting nucleus, $\Delta(\tilde
y)$, which is the solution of Eq.(34) at $\alpha = 22.5^0$, on the
coordinate $\tilde y$ (see Eq.(33)). We pay attention that
$\Delta(\tilde y)$ changes its sign in space.}
\end{figure}

To summarize, for the first time, we have suggested an adequate
mathematical apparatus to calculate the four-fold anisotropy of
the parallel upper critical magnetic field in Q2D $d$-wave
superconductors at any temperatures and numerically have
calculated its maximum value at $T=0$ in the absence of the
paramagnetic effects. We stress that our theory has been
elaborated for such Q2D superconductors [20], where $\xi_{\perp}
\geq c^*$, with $\xi_{\perp}$ being perpendicular to the
conducting layers coherent length and $c^*$ being the inter-layer
distance. This is in contrast to the so-called Lawrence-Doniach
model (see, for example, Refs.[21,22]). It is useful here to
discuss more what we meant by writing $T=0$. First of all, we have
considered a pure $d$-wave superconductor and, thus, we always
meant that $\pi T_c  \gg 1/\tau_0 \rightarrow 0$. Secondly, we
have suggested that $\pi T  \gg \omega_c(H_{c2})$, which allows to
disregard the quantum effects of electron motion in a magnetic
field [11-13]. Note that the above mentioned quantum effects have
been so far found to be important only for Q1D organic
superconductors from chemical family (TMTSF)$_2$X (X=ClO$_4,
$PF$_6$, etc.) [23].

\begin{figure}[t]
\centering
\includegraphics[width=0.5\textwidth]{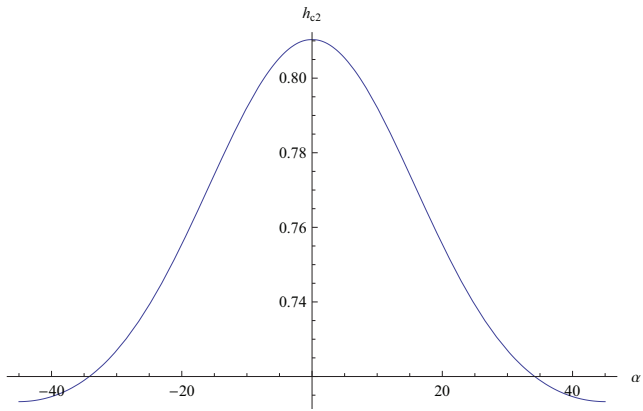}
\caption{Angular dependence of the ratio $h_{c2} (\alpha) =
H_{c2}(\alpha)/H^{GL}_{c2}(0)$, where $H_{c2}(\alpha)$ is the
parallel upper critical magnetic field for the field in-plane
direction (19) at $T=0$ and $H^{GL}_{c2}(0)$ is the
Ginzburg-Landau parallel upper critical magnetic field (31) at
$T=0$ (i.e., at $\tau$=1).}
\end{figure}

The author is thankful to N.N. Bagmet for useful discussions.

$^*$Also at: L.D. Landau Institute for Theoretical Physics, RAS, 2
Kosygina Street, Moscow 117334, Russia.

\end{document}